\documentclass[aps,twocolumn,showpacs,superscriptaddress,prl,10pt]{revtex4-1}
\usepackage{graphicx}
\usepackage{dcolumn}
\usepackage{bm}
\usepackage{color}
\usepackage{amsmath}
\usepackage{amsfonts}
\usepackage{subfigure}
\usepackage{float}
\usepackage[makeroom]{cancel}
\usepackage[section]{placeins}
\newcommand{\COMMENTED}[1]{}

\begin{document}

\author{Ettore Vitali}
\affiliation{Department of Physics, California State University Fresno, Fresno, California 93740}
\affiliation{Department of Physics, The College of William and Mary, Williamsburg, Virginia 23187}

\author{Peter Rosenberg}
\affiliation{D\'epartement de Physique \& Institut Quantique, Universit\'e de Sherbrooke, Qu\'ebec, Canada J1K 2R1}

\author{Shiwei Zhang}
\affiliation{Center for Computational Quantum Physics, Flatiron Institute, 162 5th Avenue, New York, New York 10010}
\affiliation{Department of Physics, The College of William and Mary, Williamsburg, Virginia 23187}

\title{Exotic Superfluid Phases in Spin Polarized Systems on Optical Lattices}

\begin{abstract}
Leveraging cutting-edge numerical methodologies, we study the ground state of the two-dimensional spin-polarized Fermi gas in an optical lattice. We focus on systems 
at high density and small spin polarization, corresponding to the parameter regime believed to be most favorable to the formation of the elusive Fulde-Ferrell-Larkin-Ovchinnikov (FFLO) superfluid phase. 
Our systematic study of large lattice sizes, hosting nearly $500$ atoms, provides strong evidence of the stability of the FFLO state in this regime, as well as a high-accuracy characterization of its properties.
Our results for the density correlation function reveal the existence of density order in the system, suggesting the possibility of an intricate coexistence of long-range orders in the ground state.
The ground-state properties are seen to differ significantly from the standard mean-field description, providing a compelling avenue for future theoretical and experimental explorations of 
the interplay between interaction and superfluidity in an exotic phase of matter. 
\end{abstract}

\maketitle

The relation between magnetism, superfluidity or superconductivity and other complex orders in the low temperature free energy landscape of strongly correlated quantum many-body systems has attracted tremendous interest for several decades. In conventional superconductors, 
magnetism and superconductivity are believed to compete against each other: weak magnetic fields are expelled from superconductors, while large magnetic fields destroy the superconducting order. An external magnetic field induces an alignment of the electron spins, thus disrupting the standard BCS pairing mechanism.
The BCS state is known to be energetically favorable in spin-balanced systems, however the pairing behavior in the presence of a spin polarization remains less characterized and understood.
 
The earliest theoretical proposals addressing pairing in spin polarized systems date back at least to the work of Fulde and Ferrell \cite{PhysRev.135.A550}, and Larkin and Ovchinnikov \cite{LO}, who suggested an alternative to the BCS pairing mechanism that leads to the formation of finite-momentum Cooper pairs. Despite decades of theoretical and experimental efforts, the energetic stability and precise characteristics of the FFLO state have not been conclusively determined, nor has there been any unambiguous experimental detection. The quest to observe FFLO order is an enduring challenge. 
There have been major efforts in condensed matter systems: in organic superconductors \cite{PhysRevLett.99.187002,PhysRevLett.116.067003,https://doi.org/10.1002/andp.201700282}, in the iron-based superconductor KFe$_2$As$_2$ \cite{PhysRevLett.119.217002} and in the cuprates, in connection with a possible pair-density-wave state \cite{Hamidian2016,Edkins976,Du2020,doi:10.1146/annurev-conmatphys-031119-050711}.
The rapid advances in ultracold quantum gases, which offer several significant experimental advantages,  
have opened up a new avenue both  in the continuum \cite{Radzihovsky_2010} and in optical lattices \cite{fflo_review}.
Some evidence of an FFLO state has been obtained in one-dimension \cite{Liao2010}, and intense experimental effort is underway 
in two and three dimensions. 

The underlying systems tend to be strongly interacting and the properties of the target state depend on a delicate balance of competing quantum and many-body effects. 
To date, relatively few works have addressed spin-polarized systems with calculations beyond mean-field theory,
among them dynamical mean-field theory studies \cite{dmfttroyer,PhysRevLett.101.236405,PhysRevLett.113.185301}, 
several quantum Monte Carlo calculations \cite{fflodmc,fflowolak,PhysRevB.94.075157,fflowolak},
and a recent study using field-theoretical methods \cite{Pini2021}. However, these studies are either at 
low densities \cite{PhysRevB.94.075157}, the continuum limit \cite{fflodmc,fflowolak,Pini2021}, or/and finite temperatures \cite{PhysRevLett.113.185301,fflowolak}.
Because of the scarcity of results and current limitations in computational methods (e.g., accuracy in the level of approximation, access to lower temperature or large system sizes), uncertainty remains in the nature, or even existence, of an FFLO state in optical lattice systems. There is a strong need for robust, accurate theoretical and computational results for guidance and calibration in the effort to achieve conclusive experimental detection of this elusive state.

In this work we present a set of non-perturbative numerical results that establish the stability and accurately determine the ground-state properties of
the FFLO state in the attractive two-dimensional spin-polarized Fermi gas in an optical lattice.
We concentrate on the parameter regime near half-filling with small spin polarization, which is
an ideal candidate system for the detection of an FFLO superfluid, due to the enhancement of quantum fluctuations associated with the reduced dimensionality, and the nesting effects introduced by the optical lattice at high density \cite{PhysRevA.88.043624}.
This regime has been particularly challenging for numerical many-body approaches. (The Hamiltonian can be mapped to the repulsive 
Hubbard model via particle-hole transformation, and half-filling with small spin polarization corresponds to the lightly doped $+U$ model, 
which has attracted unprecedented effort in the context of high-$T_c$ superconductivity and whose complete phase diagram remains to be determined.)
We carry out our study using the constrained path auxiliary field quantum Monte Carlo (AFQMC) technique \cite{CPMC},
reaching lattice sizes containing $\sim$\,500 atoms.
Leveraging recent methodological advances \cite{PhysRevA.100.023621}, we are able to verify the independence of the results 
from the choice of trial wave function that is used to implement the constrained path gauge condition.
This internal self-consistency, coupled with a large body of validation from prior benchmarks on the Hubbard model \cite{PhysRevX.5.041041,mingpu,stripes_science,sc_prx}, 
allow a high degree of confidence in the accuracy of these state-of-the-art many-body computations.
The results provide an ab initio characterization of the FFLO state via analysis of the bulk correlation functions;
additionally, they suggest an intriguing intertwined long-range density order. 

\begin{figure}[ht]
   \vspace{20pt}%
\includegraphics[width=\columnwidth, angle=0]{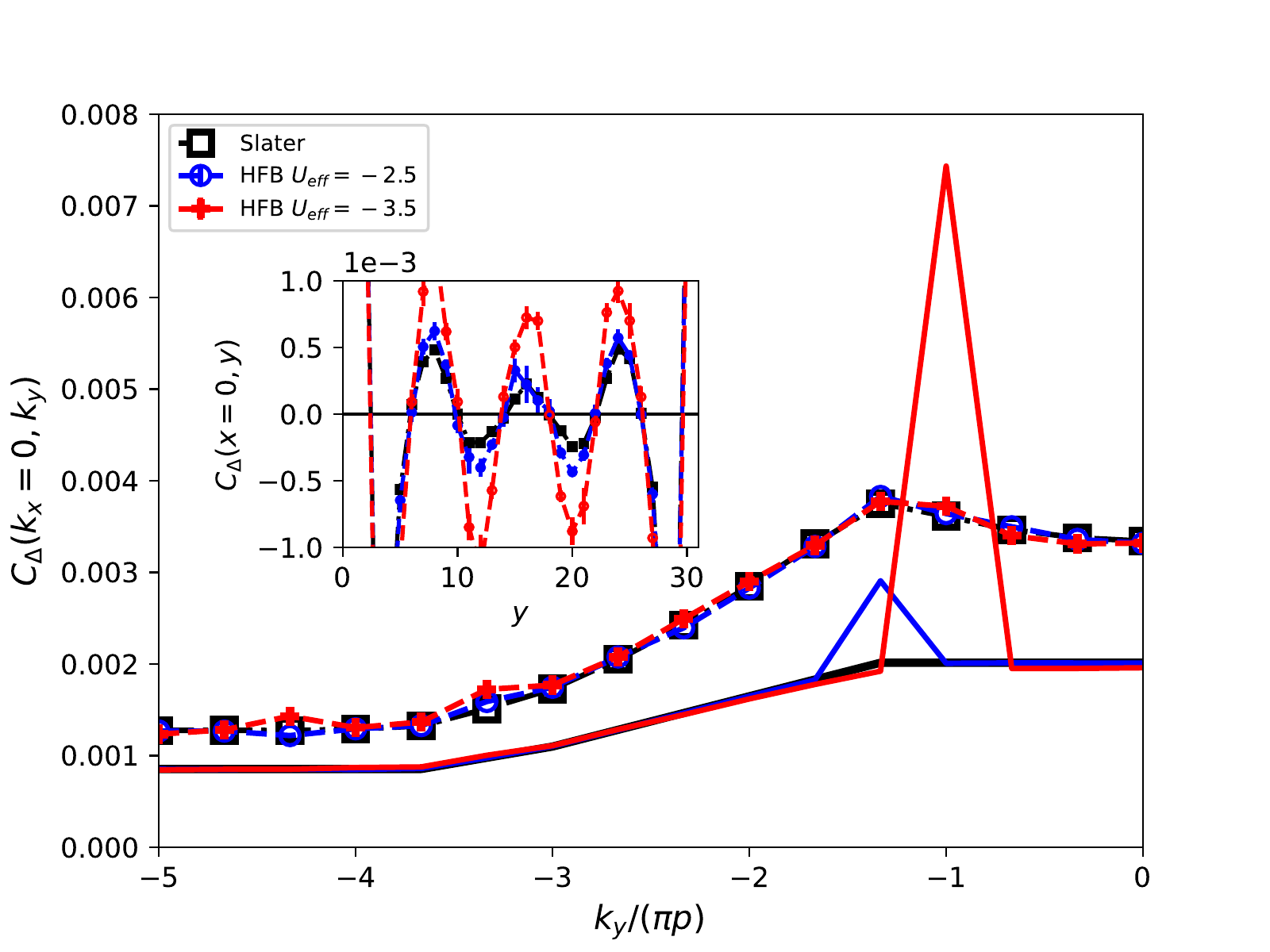}
\caption{Structure factor
of the pairing correlation for a $4\times 32$ lattice with $N_{\uparrow}=57$, $N_{\downarrow}=33$ and $U = -4$. We compare the AFQMC results (dashed lines) obtained from three different trial wave functions, the free-particle wave function (black squares), and two HFB wave functions with an effective interaction strength $U_{\rm eff}=-2.5$ (blue open circles) and $U_{\rm eff}=-3.5$ respectively (red filled circles). The wave function with $U_{\rm eff}=-2.5$ is obtained from the incommensurate HFB solution, and the wave function with $U_{\rm eff}=-3.5$ from the commensurate solution. The solid lines show the corresponding variational results.  The inset shows the corresponding modulated long-range behavior of the AFQMC results for the pairing correlation in real space. 
}
\label{fig:pairing_robustness}
\end{figure}
\vspace{1cm}

Our starting point is the two-dimensional Hubbard hamiltonian with nearest neighbor hopping, modeling a collection of $\mathcal{N}_p = N_{\uparrow} + N_{\downarrow}$ 
spin-$1/2$ fermions moving in an optical lattice:
\begin{equation}
\label{HubbardHamiltonian}
\hat{H} = -t \sum_{\langle {\bf{r}}, {\bf{r}^{\prime}} \rangle, \sigma}
 \hat{c}^{\dagger}_{{\bf{r}},\sigma} \hat{c}^{}_{{\bf{r}^{\prime}},\sigma} + U \sum_{{\bf{r}}} \hat{n}_{\uparrow}({\bf{r}})\hat{n}_{\downarrow}({\bf{r}}) 
\end{equation}
In \eqref{HubbardHamiltonian},  $U/t < 0$ is the attractive interaction strength, the label $\bf{r}$ runs over the sites of a periodic supercell 
with $\mathcal{N}_s = L_x \times L_y$  unit cells of squares with side length $a$, 
and $\hat{n}_{\sigma}({\bf{r}}) = \hat{c}^{\dagger}_{{\bf{r}},\sigma} \hat{c}^{}_{{\bf{r}}, \sigma}$ is the particle density operator with spin orientation $\sigma$  at site ${\bf{r}}$. We work in units such that $a=1$ and $t=1$.
We focus on the spin-polarized case, which can be induced by embedding the system in a Zeeman field, for example. 
In practice, we work in the sector of the Hilbert space 
with fixed values of $N_{\uparrow}$ and $N_{\downarrow}$, corresponding to a finite spin polarization $p\equiv (N_{\uparrow} - N_{\downarrow})/\mathcal{N}_s$. 
We measure the pairing correlation function $\mathcal{C}_{\Delta}({\bf{r}}) = \langle \hat{\Delta}^{\dagger}(0) \hat{\Delta}^{}({\bf{r}}) \rangle$ [where
$ \hat{\Delta}^{}({\bf{r}}) = \hat{c}^{}_{{\bf{r}},\downarrow} \hat{c}^{}_{{\bf{r}},\uparrow}$ is the on-site pairing operator and the angular brackets denote ground state averages] and its Fourier transform;
the density correlation function with the background subtracted
$\mathcal{C}_{n}({\bf{r}}) = \langle \hat{n}^{}(0) \hat{n}^{}({\bf{r}}) \rangle-\langle \hat{n} \rangle^2$;
and the spin correlation function, $\mathcal{C}_{S}({\bf{r}}) = \langle \hat{{\bf{S}}}^{}(0) \cdot \hat{{\bf{S}}}^{}({\bf{r}}) \rangle$ [where the spin density operator is defined as $ \hat{{\bf{S}}}^{}({\bf{r}}) = \sum_{\alpha,\beta} ( {\bf{\sigma}} )_{\alpha,\beta} \hat{c}^{\dagger}_{{\bf{r}},\alpha} \hat{c}^{}_{{\bf{r}}, \beta}$ with ${\bf{\sigma}}$ denoting the Pauli matrices].
To probe the physics at the bulk limit, we study large supercell sizes at given values of the density $n = (N_{\uparrow} + N_{\downarrow})/\mathcal{N}_s$ and the polarization $p > 0$, testing different aspect ratios of the supercell to detect collective modes and extrapolating the properties to  
 $N_{\sigma} \to +\infty$ and $\mathcal{N}_s \to +\infty$.

\begin{figure}[ht]
   \vspace{20pt}%
\includegraphics[width=\columnwidth, angle=0]{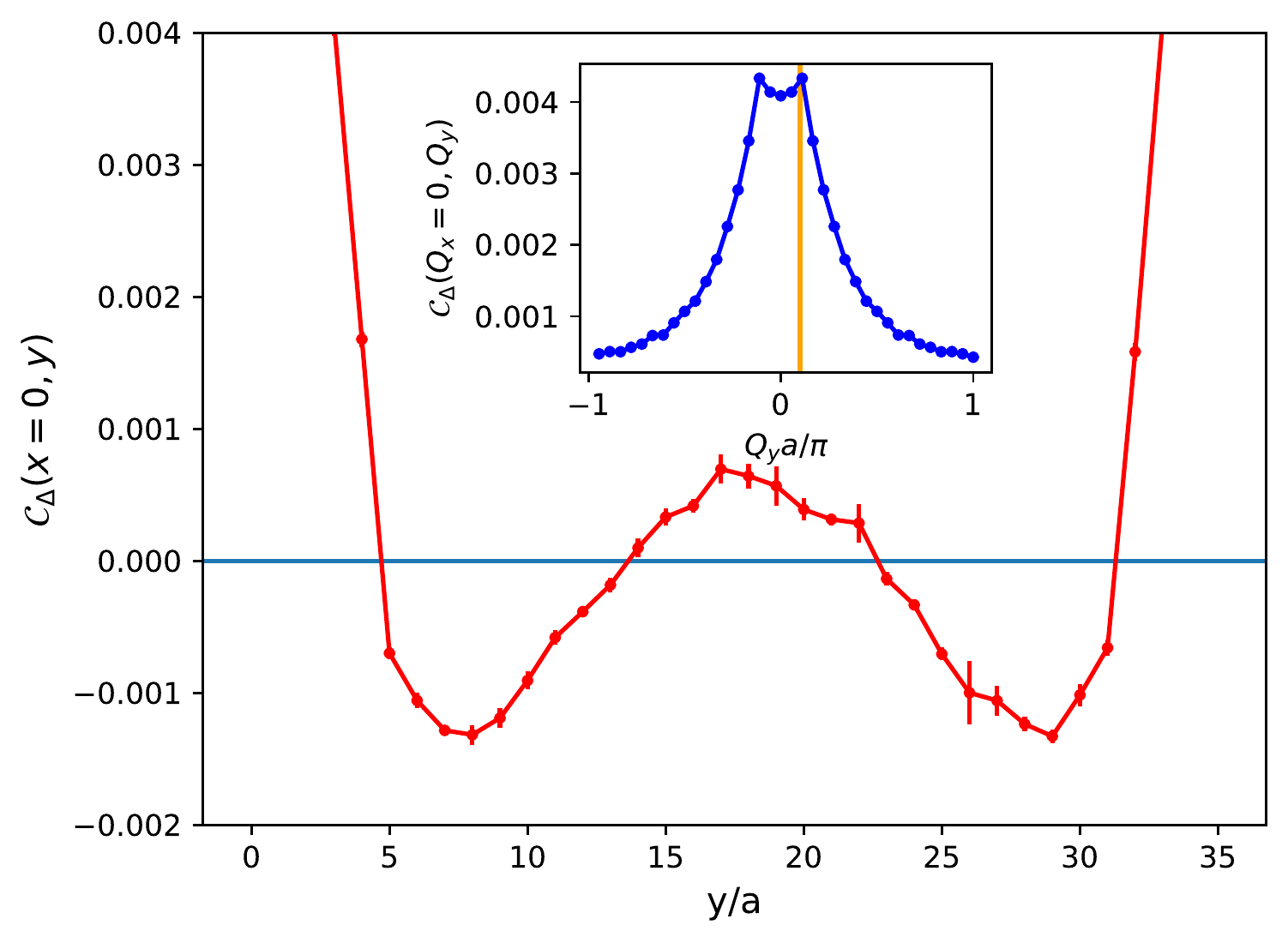}
\caption{Pairing correlation for a $4\times 36$ lattice with $N_{\uparrow}=39$, $N_{\downarrow}=29$ and $U = -4$. 
The inset shows the Fourier transform of the pairing correlation, and the vertical line shows the prediction of the pairing wave vector from \cite{PhysRevB.94.075157}.
}
\label{fig:pairing_compareDiagMC}
\end{figure}
\vspace{1cm}

Within AFQMC  \cite{CPMC}, the many-body ground-state wave function of the system $|\Psi_0\rangle$ is sampled by implementing a random walk in the manifold of Slater determinants,
driven by the imaginary-time evolution operator of the system.  
The presence of a finite polarization breaks time-reversal symmetry, which leads to the emergence of the fermion sign problem.
The constrained path gauge condition is an exact condition on the evolution of the overall sign or phase of each Slater determinant during the imaginary-time propagation.
In practice it is most commonly implemented by requiring that
each Slater determinant maintain a positive overlap with a given variational trial wave function $|\Psi_T\rangle$.
This results in a polynomially scaling algorithm but 
introduces an approximation that becomes exact if $|\Psi_T\rangle$ coincides with the ground-state wave function. 
As mentioned above, the technique has consistently performed at the highest level of accuracy with simple trial wave functions.
In order to assess and minimize the bias due to the choice of $|\Psi_T\rangle$, we systematically compare the results from 
calculations using an ensemble of trial wave functions. 
Among these is the non-interacting wave function, 
as well as fully optimized Hartree-Fock-Bogoliubov (HFB) wave functions with different values of an {\textit {effective}} interaction, $U_{\rm eff}$ (as in Ref.~\cite{mingpu} in the realm of repulsive models).
A robust FFLO order is observed even when a non-interacting trial wave function is used, and 
consistent results for the ground state correlation functions are obtained with the different choices of $|\Psi_T\rangle$, as we illustrate next.

In Fig.~\ref{fig:pairing_robustness} we present a set of calculations of the pairing correlation function $\mathcal{C}_{\Delta}({\bf{r}})$ (inset) and its Fourier transform  $\mathcal{C}_{\Delta}({\bf{k}})$ (main panel). 
The three different trial wave functions used correspond to three qualitatively different variational states, a free-fermion state and
two distinct HFB solutions: one characterized by a commensurate pairing wave vector  \cite{PhysRevA.88.043624} (i.e. ${\bf{Q}} = (0, \pi p)$), and the other by an incommensurate pairing wave vector, in this case ${\bf{Q}} = (0,  \pi p/ 0.75)$.  
In the latter two, the variational results each show a prominent peak at a different location, indicating a non-zero FFLO order parameter, $\langle \hat{c}^{}_{\downarrow}({\bf{k}}) \hat{c}^{}_{\uparrow}(-{\bf{k}}+{\bf{Q}}) \rangle$.
Despite the use of qualitatively distinct trial wave functions, AFQMC obtains a modulated real-space pairing correlation function, with 
the same wavelengths in each case, corresponding to highly consistent pairing structure factors with a 
maximum at the momentum characterizing the incommensurate HFB solution. The observation of a shallow maximum in AFQMC, instead of a prominent peak, suggests that the pairing mechanism in the many-body state is more nuanced than the simplest description of a single value for the center-of-mass momentum in the FFLO state. 

As noted earlier, we focus on the region of the density-polarization phase diagram near half-filling,
with small to moderate spin polarization. This region poses considerable challenges for numerical approaches; 
for quantum Monte Carlo techniques, the sign problem can be very severe here. Many-body calculations have thus far been away from this regime and mostly limited to fairly dilute systems. 
Before proceeding to this regime, we perform a calculation at lower density ($n=0.47$ and $p=0.07$).
 As shown in Fig.~\ref{fig:pairing_compareDiagMC}, 
 a clear long-range spatial modulation characteristic of the LO state is seen.
This result is consistent with a recent diagrammatic quantum Monte Carlo study \cite{PhysRevB.94.075157}, which found an FFLO instability in this region of the phase diagram.
The Fourier transform of the pairing correlation function, $\mathcal{C}_{\Delta}({\bf{k}})$ (inset of Fig.~\ref{fig:pairing_compareDiagMC}), 
shows a peak at wave vector ${\bf{Q}}=(0,0.1\pi) $, in very good agreement with the result from Ref.~\cite{PhysRevB.94.075157}. 
This consistency between two very different {\textit {ab initio}} methods of high accuracy is 
a strong corroboration of the FFLO order in this system.  

\begin{figure}[ht]
\begin{center}
 \vspace{0.5cm}%
\includegraphics[width=1.0\columnwidth, angle=0]{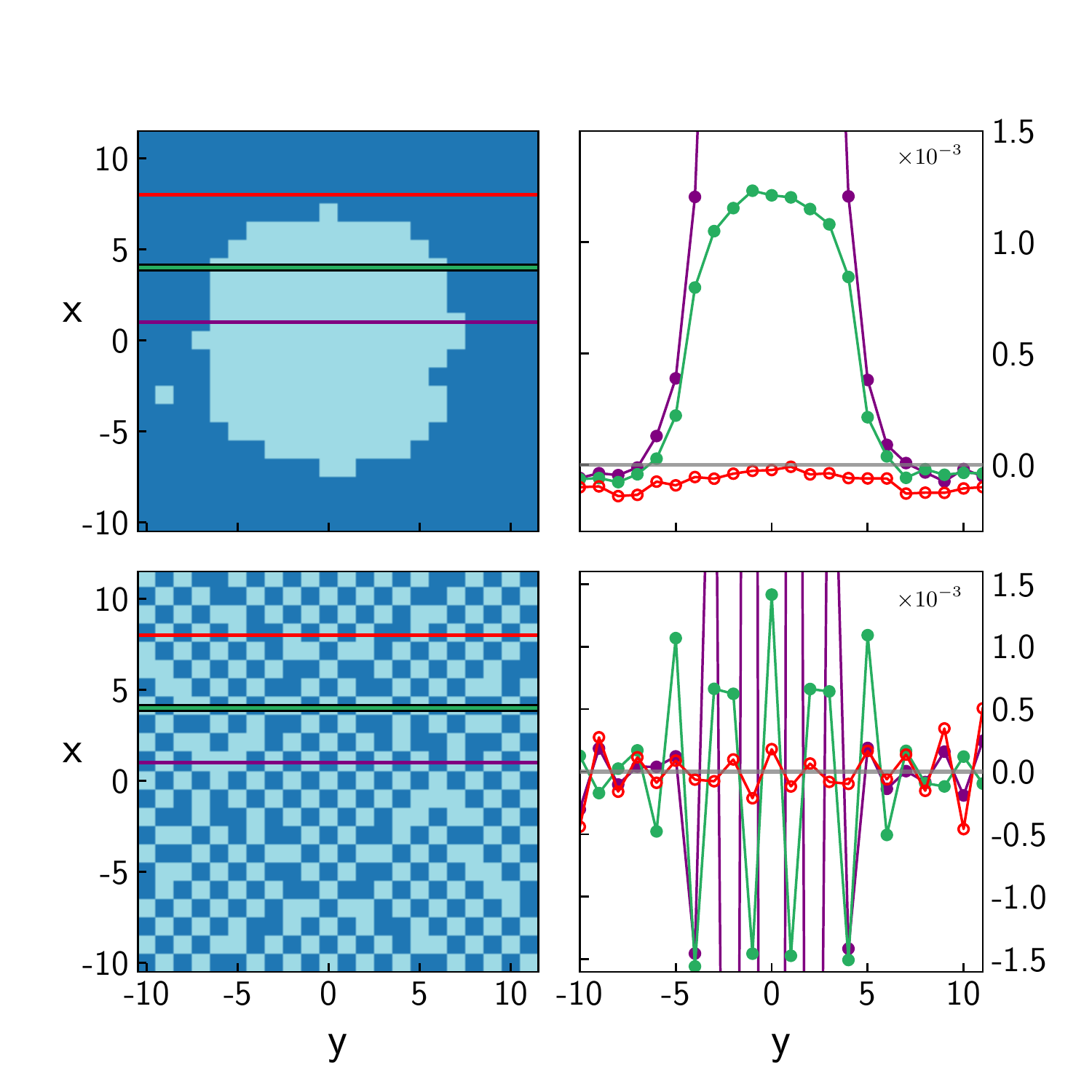}
\caption{Pairing and density correlations for a $22\times22$ system with $N_\uparrow = 217$, $N_\downarrow = 181$ and $U=-4$.  (Upper left panel) Color plot of sign$(\mathcal{C}_{\Delta}({\bf{r}}))$, to highlight the nodal structure. (Upper right panel) $\mathcal{C}_{\Delta}({\bf{r}})$ plotted along the horizontal line cuts indicated in the left panel. (Lower left panel) Color plot of sign$(\mathcal{C}_{n}({\bf{r}}))$. (Lower right panel) $\mathcal{C}_{n}({\bf{r}})$ plotted along the horizontal line cuts indicated in the left panel.
}
\label{fig:pair_pair_SD}
\end{center}
\end{figure}

We next
study more systematically the characteristics of the ground-state phases in the  high-density regime.
In Fig.~\ref{fig:pair_pair_SD} we present results for a system with $n=0.82$, $p=0.07$ and $U=-4.0$. 
We use a large supercell, consisting of $484$ sites, in order to detect long-range collective modes and minimize finite-size effects. 
The pairing correlation, shown in the upper panels, displays a clear node separating the regions with positive values of $\mathcal{C}_{\Delta}({\bf{r}})$ from those with 
negative values, and revealing a long-range modulated behavior that signals the existence of a pair-density wave in the system. 
We observe that the pair-density wave in this case is of LO form with pairing wave vectors ${\bf{Q}}$ oriented along the $x$ and $y$ directions.
(The rotational symmetry of this state can be broken with the choice of an HFB trial wave function with a single ${\bf{Q}}$ along either $x$ or $y$, which
results in a many-body ground state with pairing order along the same direction as the HFB solution.)
In this system the most prominent wave vector of the modulation is $|{\bf{Q}}| \simeq 0.09 \, \pi$;  however,
similar to the case illustrated in Fig.~\ref{fig:pairing_robustness}, the structure factor displays a broad maximum.

In the absence of spin polarization, the system is known to exhibit
a supersolid order at half-filling \cite{PhysRevA.102.053324,PhysRevLett.119.265301,PhysRevLett.62.1407,PhysRevLett.66.946},
where $s$-wave pairing coexists with a checkerboard density modulation. 
Away from half-filling, the supersolid order is believed to disappear rapidly. 
In the lower panels of Fig.~\ref{fig:pair_pair_SD}, results are shown for the density correlation function corresponding to the same polarized system.
In the long-range behavior of $\mathcal{C}_{n}({\bf{r}})$
we see a clear checkerboard density pattern, as in the supersolid phase at half-filling with no spin polarization, but with
a non-trivial modulation pattern superimposed, which appears to show phase-flip lines
roughly coincident with the location of the node of the pairing correlation function.
The spin correlations, on the other hand, are nearly uniform and do not show long-range magnetic order.
These results suggest a complex intertwined order, where the long-range pairing order appears to coexist, or compete, with an interesting 
modulated density order. The existence of these intertwined orders is likely closely connected to the pairing mechanism which, as our results indicate, may involve 
 more than a single magnitude pairing wave vector.
Further investigations are needed for better characterization and understanding.

\begin{figure}[ht]
   \vspace{20pt}%
\begin{center}
\includegraphics[width=1.\columnwidth, angle=0]{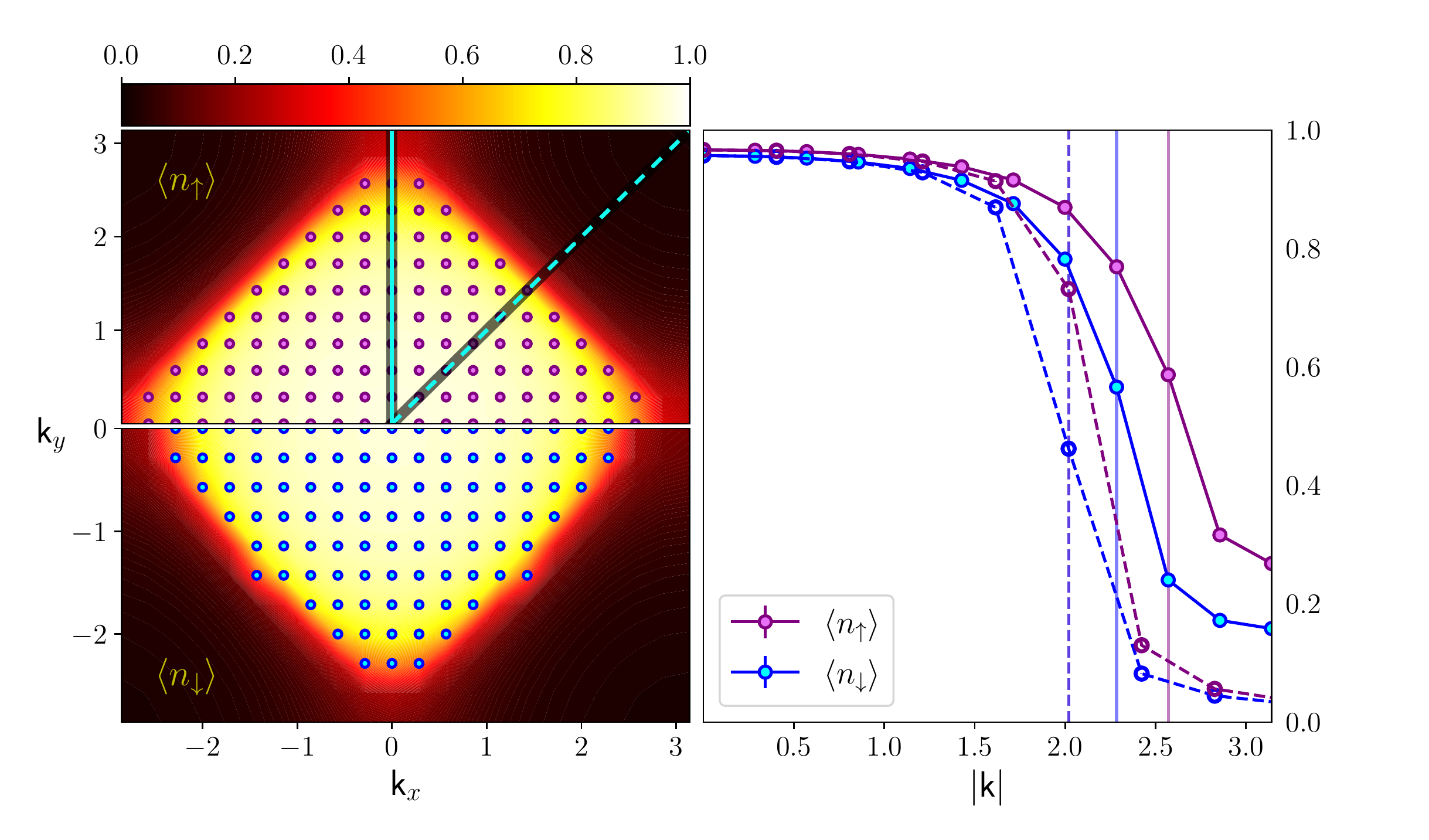}
\caption{Momentum distributions. (left) The majority (top) and minority (bottom) momentum distributions are shown by the color maps.
The circles indicate the occupied momenta in the non-interacting state.
(right) Values of $n_\uparrow({\mathbf k})$ and $n_\downarrow({\mathbf k})$ along
two cuts: $k_x=0$ (solid lines)
and the diagonal $k_x = k_y$ (dashed lines), as illustrated on the left.
The vertical lines indicate the non-interacting Fermi surfaces. 
The system is a $22\times22$ site periodic supercell with $N_\uparrow = 217$, $N_\downarrow = 181$ and $U=-4$.}
\label{fig:nk_w_slices}
\end{center}
\end{figure}
\vspace{1 cm}

To shed more light on the pairing and density order described above, we compute the spin-resolved momentum distribution of the system 
$n_{\sigma}({\bf{k}}) = \langle \hat{c}^{\dagger}_{\sigma}({\bf{k}})  \hat{c}^{}_{\sigma}({\bf{k}}) \rangle$, which is shown in Fig.~\ref{fig:nk_w_slices}.
The left panel shows color plots for both majority and minority spins,
while the right panel shows cuts of $n_{\sigma}({\bf{k}})$ along two different directions. The left panel displays also the corresponding occupied wave vectors in the non-interacting ground state, with the right panel showing the Fermi surfaces as vertical lines. From the color plot, the reorganization is evident near the Fermi surface to favor pairing with finite center-of-mass momentum in the $x$ or $y$ direction.
A relatively smooth momentum distribution is seen in the interacting system. In particular no breaches are evident in $n_{\sigma}({\bf{k}})$ 
in the flattened regions in a neighborhood of the ``node'' $(\pm \pi/2, \pm \pi/2)$, thus 
a breached pair superfluid  \cite{Liu2003} does not appear to be stable in this regime. In addition, no strong anisotropy is evident in the momentum distribution, indicating that a deformed Fermi surface superfluid state \cite{DFS1,DFS2} is likely not stable in this regime either.

In summary, we have performed extensive AFQMC 
studies
 of the ground state of the two-dimensional spin-polarized attractive Fermi gas in an optical lattice.
 We study the region of the phase diagram lightly doped from half-filling with small spin polarization, which
 is particularly challenging to treat numerically. Our results offer a beyond mean-field theory examination of the stability of the FFLO state 
 in this system and systematic characterization of its properties. We find clear signatures of the FFLO state using both a translationally and 
 rotationally invariant free-fermion trial wave function, which includes no pairing or charge order, and an ensemble of HFB wave functions, 
 which explicitly break symmetry and include pair-density waves. These high-accuracy many-body results will provide guidance and calibration 
 for experimental searches for the elusive FFLO state, in particular in the rapidly progressing field of ultra cold atoms in optical lattices.

Our results also uncover the possibility of intertwined orders in the system, comprising a non-trivial density modulation which appears to coexist, in cooperation 
or competition, with the superfluid order. Such a magnetic  superfluid phase with modulated density would be an intriguing and exotic state of matter. Experimental 
and additional theoretical studies would be invaluable to better understand its nature. Via a particle-hole transformation, our results can be mapped to the repulsive 
($+U$) Hubbard model with finite doping and spin imbalance, which continues to receive intense attention in the context of unconventional superconductivity in the cuprates.

Computing was carried out at the Extreme Science and Engineering Discovery Environment (XSEDE), which is supported by National Science Foundation grant number ACI-1053575.
The Flatiron Institute is a division of the Simons Foundation.

\end{document}